\documentclass[prd,showpcs,amsmath,amssymb,nofootinbib,longbibliography,twocolumn,showpacs,notitlepage]{revtex4-1}
\usepackage{multirow}
\usepackage{epsfig}
\usepackage{amsmath}
\usepackage{bm}
\usepackage{times}
\usepackage{graphicx}
\usepackage{color}
\usepackage{slashed}
\usepackage{graphicx} 
\usepackage{amsmath}
\usepackage[latin1]{inputenc}
\usepackage{hyperref}
\usepackage{soul}
\usepackage{lipsum}

\def\bea{\begin{eqnarray}}
\def\eea{\end{eqnarray}}
\def\bean{\begin{equation*}}
\def\eean{\end{equation*}} 
\def\nn{\nonumber}
\def\beaal{\begin{align}}
\def\eeaal{\end{align}}

\begin{document}
 
\title{Gravitational Wave Signatures of Lepton Universality Violation}

\author{Bartosz~Fornal}
\affiliation{\vspace{1mm}Department of Physics and Astronomy, University of Utah, Salt Lake City, UT 84112, USA\vspace{1mm}}
\date{\today}

\begin{abstract}
We analyze the prospects for using  gravitational waves produced in early universe phase transitions  as a complementary probe of the flavor anomalies in $B$
 meson decays. We focus on the Left-Right SU(4) Model, for which  the  strength of the observed lepton universality violation and consistency with other experiments impose a vast hierarchy between the symmetry breaking scales. This leads to a multipeaked gravitational wave signature within the reach of upcoming gravitational wave detectors.
\vspace{8mm}
\end{abstract}

\maketitle

\section{Introduction}\label{1} 
Although the Standard Model (SM) does not provide all the answers to fundamental questions in particle physics and needs to be augmented by new physics, nearing  half a century since its formulation \cite{Glashow:1961tr,Weinberg:1967tq,Salam:1968rm,SU(3),Fritzsch:1973pi} it has certainly stood the test of time with respect to its predictive power.
A huge number of models beyond the SM have been constructed proposing   solutions to the outstanding problems, however, it is not certain which of them, if any, is realized in nature. At this time, guidance from experiment is especially important in order to achieve further progress on the theory side. 
\vspace{2mm}

So far, among the strongest experimental hints of  new physics are the indications of lepton universality violation in $B$ meson decays, the so-called $R_{K^{(*)}}$ and $R_{D^{(*)}}$ anomalies.
Although the $R_{D^{(*)}}$ anomalies (reported by BaBar \cite{Lees:2013uzd}, Belle \cite{Sato:2016svk} and LHCb \cite{Aaij:2017uff}) have not been confirmed in the most recent set of Belle data \cite{Abdesselam:2019dgh}, and the $R_{K^{*}}$ anomaly reported by LHCb \cite{Aaij:2017vbb} has become less significant  \cite{Abdesselam:2019wac}, 
the $R_{K}$ anomaly  \cite{Aaij:2014ora}  has persisted  with new  LHCb data \cite{Aaij:2019wad}. 
\vspace{2mm}
 
From an effective theory point of view  \cite{Kosnik:2012dj,Alonso:2014csa,Bhattacharya:2014wla,Alonso:2015sja,Bhattacharya:2016mcc,Geng:2017svp,Alok:2017jaf,Buttazzo:2017ixm,Kumar:2018kmr} the observed signals of lepton universality violation are best accounted for  by either the vector leptoquark $(3,1)_{\frac23}$ or  $(3,3)_{\frac23}$.
A natural origin of the former in the context of  flavor anomalies has been proposed in \cite{Assad:2017iib}, where it was suggested that this leptoquark can be the gauge boson of a Pati-Salam-type unified model. This has been followed by several efforts aimed at explaining either both the $R_{K^{(*)}}$ and $R_{D^{(*)}}$ anomalies \cite{Calibbi:2017qbu,DiLuzio:2017vat,Bordone:2017bld,Barbieri:2017tuq,Blanke:2018sro,Greljo:2018tuh,DiLuzio:2018zxy,Fuentes-Martin:2019bue,Fuentes-Martin:2020bnh,Guadagnoli:2020tlx} or just the $R_{K^{(*)}}$ \cite{new,Fornal:2018dqn} through the  vector leptoquark $(3,1)_{\frac23}$ in UV complete models. Other explanations of the  anomalies include  $Z'$ bosons \cite{Cline:2017lvv,Marzocca:2018wcf,Guadagnoli:2018ojc,Li:2018rax,Faber:2018qon,Allanach:2018lvl}  and  scalar leptoquarks \cite{Cai:2017wry,Heeck:2018ntp,Becirevic:2018afm,Bigaran:2019bqv}. 
\vspace{2mm}

Since the latest  results on $R_{D^{(*)}}$ from Belle  \cite{Abdesselam:2019dgh}  are consistent with the SM, lowering the overall significance of those anomalies, in this paper we focus on the solution to just the  $R_{K^{(*)}}$ anomalies offered by the \emph{Left-Right SU(4) Model} \cite{Fornal:2018dqn}. This is the only model for the flavor anomalies proposed  so far which does not require any mixing between quarks and new vector-like fermions. 
Apart from the existing experimental searches for lepton universality violation, the only other conventional way of looking for signatures of this model is to produce the vector leptoquark in particle colliders. However, given its large mass of $\sim 10 \ {\rm TeV}$, this would  require using  the $100 \ {\rm TeV}$ Future Circular Collider, whose construction has not yet  been approved.

A new window of opportunities for probing particle physics models has recently been opened by gravitational wave experiments.
The gravitational wave detectors LIGO \cite{TheLIGOScientific:2014jea} and Virgo \cite{TheVirgo:2014hva}, in addition to observing signals from astrophysical phenomena such as black hole and neutron star mergers, have unique capabilities of detecting  the imprints of cosmic  events in the early universe, providing access to regions of parameter space unexplored so far in various extensions of the SM. This will be even more promising with future experiments such as the Laser Interferometer Space Antenna (LISA) \cite{Audley:2017drz}, Cosmic Explorer (CE) \cite{Reitze:2019iox}, Einstein Telescope (ET) \cite{Punturo:2010zz}, DECIGO \cite{Kawamura:2011zz} or Big Bang Observer (BBO) \cite{Crowder:2005nr}. 

One class of particle physics signals that gravitational wave detectors are sensitive to arises from early universe phase
transitions. If the scalar potential has a nontrivial vacuum structure, the universe could have settled in a state
which, as the temperature dropped, became metastable. The universe would then undergo a transition via thermal fluctuations 
from the false vacuum to the true vacuum.  During such a first order phase transition, bubbles of true vacuum would form in different patches of the universe and start expanding. Gravitational
waves would be generated from bubble wall collisions, magnetohydrodynamic turbulence and sound shock waves of the
early universe plasma generated by the bubble's violent expansion. 
At the Lagrangian level of a theory, a phase transition is triggered by spontaneous symmetry breaking.  In models with a rich gauge structure, multiple steps of symmetry breaking can occur, resulting in a chain of phase transitions, each  generating gravitational waves.

First order phase transitions from  symmetry breaking have been studied with respect to their predictions regarding the gravitational wave signals  in various models of new physics (see, e.g., \cite{Apreda:2001us,Grojean:2006bp,Leitao:2012tx,Schwaller:2015tja,Huang:2017laj,Huang:2017rzf,Demidov:2017lzf,Hashino:2018zsi,Madge:2018gfl,Ahriche:2018rao,Brdar:2018num,Croon:2018kqn,Angelescu:2018dkk,Hasegawa:2019amx,Dev:2019njv,Brdar:2019fur,Wang:2019pet,Greljo:2019xan,vonHarling:2019gme,Hall:2019ank,Huang:2020bbe}). 
Here we investigate the complementarity between gravitational wave experiments and direct searches for lepton universality violation. Such a connection has recently been made in \cite{Greljo:2019xan} in the context of the Pati-Salam Cubed Model \cite{Bordone:2017bld}, which consists of three copies of the Pati-Salam gauge group, each for a different family of particles. 
In the Left-Right SU(4) Model which we are considering, the gauge group is common to all the families. The symmetry breaking pattern consists of three steps, each leading to a distinct peak in the gravitational wave spectrum. The position of the two lower-frequency peaks in the three-peaked gravitational wave spectrum is determined by the magnitude of the flavor anomalies, offering a  way to discriminate the model.

\section{Left-right SU(4) model}

In this section we provide a summary of the most important properties of the model
(for further details, see \cite{Fornal:2018dqn}). 
\vspace{1mm}

\begin{table}[htbp] 
\begin{center}
\begingroup
\setlength{\tabcolsep}{6pt} 
\renewcommand{\arraystretch}{1.5} 
\begin{tabular}{  |c | c |c|} 
\hline
\multicolumn{3}{|c|}{Fermion fields}\\
\hline
$\mathcal{G}$ & SM rep. & masses  \\ 
\hline
\hline
 \raisebox{-1.5ex}[0pt]{$\hat\Psi_{L} =(4,1,2, 0)$}  & $Q_L = (3,2)_{\frac16}$  & \\ [2pt]
  &  \raisebox{1.5ex}[0pt]{$L_L = (1,2)_{-\frac12}$}  &\\ [-4pt]
  \cline{1-2}
 \raisebox{-1.5ex}[0pt]{$\hat\Psi_{R}^d=(1,4,1, -\tfrac12)$} & $d_R = (3,1)_{-\frac13}$  & \raisebox{3.5ex}[0pt]{SM masses for} \\ [0pt]
 & \raisebox{1.5ex}[0pt]{$e_R=  (1,1)_{-1}$}  & \raisebox{5.5ex}[0pt]{quarks and leptons}\\ [-6pt]
\cline{1-2}
  \raisebox{-1.5ex}[0pt]{$\hat\Psi_{R}^u=(1,4,1, \tfrac12)$}  & $u_R = (3,1)_{\frac23}$  & \raisebox{3.0ex}[0pt]{$m_{\nu} \sim \frac{v^2}{v_{10}}$} \\ [2pt]
  &  \raisebox{1.5ex}[0pt]{$\nu_R = (1,1)_{0}$}  & \raisebox{4.0ex}[0pt]{$M_{\nu_R} \sim v_{10}$} \\ [-6pt]
\hline
 \raisebox{-1.5ex}[0pt]{$\hat\chi_{L}=(\bar4,1,2, 0)$} & $Q'_L = (\bar3,2)_{-\frac16}$ &  \\ [2pt]
 &  \raisebox{1.5ex}[0pt]{$L'_L =  (1,2)_{\frac12}$} &   \raisebox{1.5ex}[0pt]{$(M_{Q'})_{ij} = \frac{Y_{ij}}{\sqrt2}v_\Sigma$} \\ [-4pt]
\cline{1-2}
 \raisebox{-1.5ex}[0pt]{$\hat\chi_{R}=(1,\bar4,2, 0)$} & $Q'_R = (\bar3,2)_{-\frac16}$ &  \raisebox{-0.5ex}[0pt]{$(M_{L'})_{ij} = \frac{Y_{ij}}{\sqrt2}z \,v_\Sigma$} \\ [-3.5pt]
&  \raisebox{0.5ex}[0pt]{$L'_R =  (1,2)_{\frac12}$} &  \raisebox{0.5ex}[0pt]{} \\
 \hline
\end{tabular}
\vspace{5mm}

\begin{tabular}{  |c | c |c|} 
\hline
\multicolumn{3}{|c|}{Scalar fields}\\
\hline
$\mathcal{G}$ & SM rep. & masses \\ [2pt]
\hline
\hline
 \raisebox{-1.7ex}[0pt]{$\hat\Sigma_R = (1, 4, 1, \tfrac12)$} &  \raisebox{-1.7ex}[0pt]{$(3,1)_{\frac23}$\,, \ $(1,1)_{0}$}  & \!\!radial mode\!\!  \\[2pt]
 & & \raisebox{2.5ex}[0pt]{$\sqrt{2\lambda_R} \, v_R\!$}\\[-8pt]
\hline
 \raisebox{-1.7ex}[0pt]{$\hat\Sigma_L = (4, 1, 1, \tfrac12)$} &  \raisebox{-1.7ex}[0pt]{$(3,1)_{\frac23}$\,, \ $(1,1)_{0}$}  & \!\!radial mode\!\!  \\[2pt]
 & & \raisebox{2.5ex}[0pt]{$ \ \  \sqrt{2\lambda_L} \, v_L\! \ \  $}\\[-8pt]
\hline
 & $(8,1)_0$\,,  $(3,1)_{\frac23}$\,,  $(1,1)_{0}$\,, &  \raisebox{0ex}[0pt]{\!\!radial modes\!\!} \\[-2pt]
 \raisebox{1.7ex}[0pt]{$\hat\Sigma = (\bar4, 4, 1, 0)$} &    \raisebox{0.2ex}[0pt]{$(\bar3,1)_{\text{--}\frac23}$\,, $(1,1)_0$}& \!$\sim \sqrt{ \lambda^{(\prime)}_\Sigma} \,v_\Sigma$\!\\[2pt]
\hline
 & \!\! $(8, 2)_{\frac12}$\,,  $(3, 2)_{\frac76}$\,,  $ (1,2)_{\frac12}$\,,\!\! &  \raisebox{-0.1ex}[0pt]{$\sim M$} \\[-2pt]
 \raisebox{1.7ex}[0pt]{$\hat{H}_{d}=(4, \bar4, 2, \tfrac{1}{2})$}   & $(\bar3,2)_{\text{--}\frac16}$\,,  $(1,2)_{\frac12}$& $\!m_{S'_1}  = m_{h}\!$  \\[3pt]
\hline
& \!\!\! $(8, 2)_{\text{--}\frac12}$, $(3, 2)_{\frac16}$, $(1,2)_{\text{--}\frac12}$,\!\!\!& \\[-2pt]
 \raisebox{1.7ex}[0pt]{$\!\hat{H}_{u} = (4, \bar4, 2, \text{--}\tfrac{1}{2})$\!}& $(\bar3,2)_{\text{--}\frac76}$\,,  $(1,2)_{\text{--}\frac12}$ &   \raisebox{1.7ex}[0pt]{$\sim M$}\\[3pt]
\hline  
\!$\hat\Phi_{10} \!=\!(1,\overline{10},1,\text{--}1)$\!& \!\!\!$(\bar6,1)_{\text{--}\frac43}$\,, $(\bar3,1)_{\text{--}\frac23}$\,, $(1,1)_{0}$\!\!\!& $\sim \,v_{10}$ \\[3pt]
\hline
\end{tabular}
\vspace{5mm}

\begin{tabular}{  |c |c |c|} 
\hline
\multicolumn{3}{|c|}{Vector fields}\\
\hline
 $\mathcal{G}$ & SM rep. & masses \\ 
\hline
\hline
& \raisebox{-0.2ex}[0pt]{$g$, $W^a_\mu$, $Z$} &  \raisebox{-0.2ex}[0pt]{SM particles} \\[2pt]
\cline{2-3}
 \raisebox{-1.5ex}[0pt]{$G^A_{R\mu}$} & \raisebox{-0.0ex}[0pt]{$X_R = (3,1)_{\frac23}$} &  \\[4pt]
 & \raisebox{1.3ex}[0pt]{$X_R^\dagger = (\bar3,1)_{-\frac23}$} & \raisebox{3.0ex}[0pt]{$M_{X_R} =\frac12 g_R v_R$} \\[-3pt]
\cline{2-3}
  \raisebox{-1ex}[0pt]{$Z'_\mu$} &  \raisebox{-1ex}[0pt]{$Z'_R = (1,1)_0$} & \raisebox{-1ex}[0pt]{$M_{Z'_R} = \frac12 \sqrt{{g'}_{\!\!1}^2+\frac32 g_R^2}\ v_R$} \\[8pt]
  \cline{2-3}
 & \raisebox{-0.0ex}[0pt]{$X_L = (3,1)_{\frac23}$} &  \\[4pt]
 \raisebox{2.1ex}[0pt]{$G^A_{L\mu}$} & \raisebox{1.3ex}[0pt]{$X_L^\dagger = (\bar3,1)_{-\frac23}$} & \raisebox{3.0ex}[0pt]{$M_{X_L} =\frac12 g_L v_L$} \\[-3pt]
\cline{2-3}
 &  \raisebox{-1.6ex}[0pt]{$Z'_L = (1,1)_0$} & \raisebox{-1.8ex}[0pt]{$\!M_{Z'_L} = \sqrt{\frac{ 3{g'_1}^{\!2}(g_L^2+g_R^2)+\frac92\,g_L^2g_R^2}{8({g'_1}^{\!2}\!+\frac32g_R^2)}} \ v_L$\!} \\[14pt] 
 \cline{2-3}
 \raisebox{4ex}[0pt]{$W^a_\mu$}&\raisebox{-0.3ex}[0pt]{$G' = (8,1)_0$} & \raisebox{-0.3ex}[0pt]{$M_{G'} =\frac1{\sqrt2}\sqrt{g_L^2+g_R^2}\ v_\Sigma$} \\[4pt]
\hline
\end{tabular}
\endgroup
\end{center}
\vspace{-1.5mm}
\caption{The fermion, scalar and vector particle content of the model. The masses  were calculated assuming  the hierarchical vev structure  $v_{10} \gg M \gg v_R \gg v_L \gg v_\Sigma$.}\vspace{0mm}
\label{table11}
\end{table}

The model is based on the gauge group
\bea
\mathcal{G} = {\rm SU}(4)_L \times {\rm SU}(4)_R \times {\rm SU}(2)_L \times {\rm U}(1)' \ .
\eea
The fermion, scalar and vector particle contents  are provided in Table \ref{table11}.
The gauge group $\mathcal{G}$ is broken by the vacuum expectation values (vevs) of the scalar fields $\hat\Sigma_R$, $\hat\Sigma_L$ and $\hat\Sigma$. The parameters of the scalar potential can be chosen such that  the following vev structure is obtained, 
 \bea\label{vev}
\langle \hat\Sigma_R^i\rangle = \frac{v_R}{\sqrt2}\,\delta^{i4} , \ \ \ \ \langle \hat\Sigma_L^i\rangle = \frac{v_L}{\sqrt2}\,\delta^{i4} , \nn\\[3pt]
\langle \hat\Sigma\rangle = \frac{v_\Sigma}{\sqrt2}\,{\rm diag}(1,1,1,z) \ , \hspace{7mm}
\eea
where $z>0$. The ${\rm SU}(4)_R$ symmetry is broken at a high scale $v_R$ in order to suppress right-handed lepton flavor changing currents and comply with the stringent experimental bounds, whereas the other scales, $v_L$ and $v_\Sigma$, are constrained by the size of the $R_{K^{(*)}}$ anomalies to be  much lower than $v_R$. We make an additional assumption that there is also a hierarchy between the scales $v_L$ and $v_\Sigma$, i.e.,
\bea\label{vevst}
v_R \gg v_L \gg v_\Sigma \ .
\eea
This implies the following symmetry breaking pattern (with the numerical choice for the vevs explained below):\vspace{2mm}
 \begin{equation}
     \begin{array}{c}
{\rm SU}(4)_L \times {\rm SU}(4)_R \times {\rm SU}(2)_L \times {\rm U}(1)' \nn\\[3pt]
\hspace{14mm}\bigg\downarrow \hspace{3mm}{\scriptstyle v_R \ \sim \ 5000 \ {\rm  TeV}}\nn\\[12pt]
{\rm SU}(4)_L \times {\rm SU}(3)_R \times {\rm SU}(2)_L \times {\rm U}(1)''\nn\\[3pt]
\hspace{11.2mm}\bigg\downarrow \hspace{3mm}{\scriptstyle v_L \ \sim \ 40 \ {\rm  TeV}}\nn\\[12pt]
{\rm SU}(3)_L \times {\rm SU}(3)_R \times {\rm SU}(2)_L \times {\rm U}(1)_Y\nn\\[3pt]
\hspace{10mm}\bigg\downarrow \hspace{3mm}{\scriptstyle v_\Sigma \ \sim \ 7 \ {\rm  TeV}}\nn\\[12pt]
 {\rm SU}(3)_c \times {\rm SU}(2)_L \times {\rm U}(1)_Y\\[7pt]
     \end{array}
   \end{equation}
The ${\rm U}(1)'$ charge $Y'$, the ${\rm U}(1)''$ charge $Y''$ and the SM hypercharge $Y$ are related via
\bea
&&Y'' = Y' + \tfrac{1}{6}\,{\rm diag}(1,1,1,-3)\ ,\nn\\
&&Y = \,Y'' +   \tfrac{1}{6}\,{\rm diag}(1,1,1,-3) \ .
\eea
The covariant derivative can be written as
\bea
\begin{aligned}
D_\mu =   \partial_\mu &+ i g_L \hspace{0.4mm}G_{L\mu}^A T_L^A +  i g_R \hspace{0.4mm}G_{R\mu}^A T_R^A \\
&+ i g_2 \,W_\mu^a \,T^a + i g_{1}' \,Y'_{\mu}\, Y' \ ,
\end{aligned}
\eea
where  the index $A = 1, ..., 15$, the index $a=1,2,3$, and\break $T_L^A$, $T_R^A$, $T^a$, $Y'$ are the ${\rm SU}(4)_L$, ${\rm SU}(4)_R$, ${\rm SU}(2)_L$, ${\rm U}(1)'$ generators, respectively. At the low scale, the gauge couplings $g_L$, $g_R$, $g'_1$ are related to the SM gauge couplings $g_s$, $g_1$ via
\bea\label{six}
g_s =\! \frac{g_L g_R}{\sqrt{g_L^2 + g_R^2}}\ , \ \ \ g_1 \!=\! \frac{g'_1g_L g_R}{\sqrt{\tfrac23{g'_1}^{\!2}(g_L^2+g_R^2)+g_L^2 g_R^2}} \ . \ \ \ \ \ \ \ 
\eea

The Lagrangian terms describing  the fermion masses are
\bea
\mathcal{L}_{f} &=&  
\big[\,y_{ij}^{d}\,\overline{\hat\Psi^{i}_L} \hat{H}_d \hat\Psi_{R}^{d\hspace{0.2mm}j}   +y_{ij}^{u}\,\overline{\hat\Psi^i_L} \hat{H}_u \hat\Psi_{R}^{u j}+ Y_{ij}\,\overline{\hat\chi^{i}_L} \,\hat{\Sigma} \,\hat\chi_{R}^{j}\,\big] \nn\\
 &+&{\rm h.c.} +  y_{ij}^{u\hspace{0.2mm}\prime}\, \overline{(\hat\Psi_{R}^{u i})^{c}}\,\hat{\Phi}_{10}\hat\Psi_{R}^{uj}\ ,
\eea
where the scalar field $\hat\Phi_{10} = (1,\overline{10},1,-1)$ develops a high-scale vev $v_{10}\sim 10^{13} \ {\rm GeV}$ and provides a seesaw mechanism for the neutrino masses, $m_\nu \sim v^2/v_{10}$ with $v$ being the SM Higgs vev. After symmetry breaking down to the SM gauge group, the fermion mass terms become
\bea
\mathcal{L}_{f}&\supset&  \Big[\,y_{ij}^{d}\,\overline{L^i_L} {S}_1  e_{R}^{j} + y_{ij}^{d}\,\overline{Q^i_L} {S}_2 d_{R}^{j}  + y_{ij}^{u}\,\overline{L^i_L} {S}_3^*  \nu_{R}^{j} \nn\\
&+& y_{ij}^{u}\,\overline{Q^i_L} {S}_4^*  u_{R}^{j}   + \tfrac{1}{\sqrt2}Y_{ij}v_\Sigma\big(\overline{{Q'}^{i}_{\!\!L}} \, {Q'}_{\!\!R}^{j} + z\,\overline{{L'}^{i}_{\!\!L}} \, {L'}_{\!\!R}^{j}\big)\,\Big] \nn\\
&+& {\rm h.c.} + y_{ij}^{u\hspace{0.2mm}\prime}\, v_{10} \,\overline{(\nu_R^{i})^c}\,\nu_R^j\ .
\eea

The scalar sector is described by the Lagrangian
\bea
\mathcal{L}_{s} &=&  
 |D_\mu \hat\Sigma_R|^2\!+\! |D_\mu \hat\Sigma_L|^2  \!+\! |D_\mu \hat\Sigma|^2 \!+\!|D_\mu \hat{H}_d|^2 \!+\! |D_\mu \hat{H}_u|^2  \nn\\
 &+&  |D_\mu \hat{\Phi}_{10}|^2+  V(\hat\Sigma_R, \hat\Sigma_L, \hat\Sigma, \hat{H}_d, \hat{H}_u, \hat\Phi_{10}) \ ,
\eea
where the scalar potential contains the following terms\break (traces are implicit),
\bea\label{scPot}
V \supset &-& \mu_{R}^2\, |\hat\Sigma_R|^2+ \lambda_{R} |\hat\Sigma_R|^4  -   \mu_{L}^2\, |\hat\Sigma_L|^2+ \lambda_{L} |\hat\Sigma_L|^4 \nn\\
&-& \mu_{\Sigma}^2\, |\hat\Sigma|^2+\lambda_\Sigma (\hat\Sigma\hat\Sigma^\dagger)^2 + \lambda_{\Sigma}' |\hat\Sigma\hat\Sigma^\dagger|^2 \nn\\
&+& \lambda_{12} |\hat\Sigma_L|^2 |\hat\Sigma_R|^2 +  \lambda_{13} |\hat\Sigma_L|^2|\hat\Sigma|^2 + \lambda_{23}  |\hat\Sigma_R|^2 |\hat\Sigma|^2\nn\\
&+& \lambda'_{13}|\hat\Sigma_L \hat\Sigma|^2 +  \lambda'_{23}|\hat\Sigma_R^\dagger \hat\Sigma|^2 + \big[\kappa \,\hat\Sigma_L \hat\Sigma \,\hat\Sigma_R^\dagger + {\rm h.c.}\big]\nn\\
&+& M_u^2 |\hat{H}_u|^2 + M_d^2 |\hat{H}_d|^2 \ .
\eea
In Eq.\,(\ref{scPot}) we left out the cross terms between $\hat\Sigma_R$, $\hat\Sigma_L$, $\hat\Sigma$ and $\hat{H}_{u}$, $\hat{H}_{d}$;  the full form of the scalar potential is given in \cite{Fornal:2018dqn}.

As discussed in \cite{Fornal:2018dqn}, it is possible to tune the parameters of the scalar potential  such that only one linear combination of the fields $S_1$, $S_2$, $S_3$, $S_4$ is light, reproducing the SM scalar sector at low energies ($S'_1 \equiv H$). The remaining  fields $S'_{2,3,4}$ and all other components of $\hat{H}_u$ and $\hat{H}_d$ have masses set by the hard mass parameters $M_u$, $M_d$, which we take to be $M_u = M_d \equiv M\gg v_R$. The relative mass hierarchies between the SM down-type quarks and charged leptons are reproduced reasonably well within this minimal setup. One can also introduce into the model the scalar representation $\hat\Phi_{15}=(15,1,1,0)$ that develops a vev at a high scale and leads to terms $\overline{\hat\Psi^i_{L}}\hat{H}_d \hat{\Psi}_R^{d \,j}\hat{\Phi}_{15}/\Lambda$ providing distinct contributions to the quark and  lepton masses.
\vspace{1mm}

The  Lagrangian terms involving the fermion and vector fields are given by
\bea
\mathcal{L}_{v}&=& \overline{\hat\Psi}_{L} i\slashed{D}\, {\hat\Psi_{L}} + \overline{\hat\Psi} {}^u_{R}\, i\slashed{D} \,{\hat\Psi_{R}^u} + \overline{\hat\Psi} {}_{R}^d \,i \slashed{D} \,{\hat\Psi_{R}^d}\nn\\
& +& \overline{\hat\chi} {}_{R} \,i \slashed{D} \,{\hat\chi_{R}}  + \overline{\hat\chi} {}_{L} \,i \slashed{D} \,{\hat\chi_{L}}  \ ,
\eea
which, at the low scale, result in the following interactions between quarks, leptons and gauge leptoquarks,
\bea
\mathcal{L}_{v} &\supset&   \frac{g_L}{\sqrt2}\, X_{L \mu} \Big[L^u_{ij}\,(\bar u^i\gamma^\mu P_L\,\nu^j)+L^d_{ij}\,(\bar d^i \gamma^\mu P_L \,e^j) \Big]\\
&+& \frac{g_R}{\sqrt2}\, X_{R\mu} \Big[R^u_{ij}\,(\bar u^i\gamma^\mu P_R\,\nu^j)+R^d_{ij}\,(\bar d^i \gamma^\mu P_R\, e^j) \Big]  \!+ {\rm h.c.} , \nn
\eea
where $L^u$, $L^d$, $R^u$ and $R^d$ are mixing matrices. They are all unitary and related to the Cabibbo-Kobayashi-Maskawa matrix and the Pontecorvo-Maki-Nakagawa-Sakata matrix via $L^u = V_{\rm CKM}L^d U_{\rm PMNS}\,$ and $\,R^u = V_{\rm CKM}R^d U_{\rm PMNS}$. \vspace{1mm}

To circumvent the stringent experimental constraints on lepton universality violation \cite{Britton:1992pg,Britton:1993cj,Czapek:1993kc,Ambrose:1998cc,Ambrose:1998us,Ambrose:2000gj,Appel:2000tc,Sher:2005sp,Ambrosino:2009aa,Aubert:2006vb,Aubert:2007mm,Aubert:2007rn,Aubert:2008cu,Aaltonen:2009vr,Aaij:2017cza,Aaij:2017vad,Aaij:2017xqt,Aubert:2006cz,Miyazaki:2007jp,Miyazaki:2010qb,Bertl:2006up},  the scale of ${\rm SU}(4)_R$ breaking needs to be $v_R \gtrsim 5000 \ {\rm TeV}$ for a generic unitary  matrix $R^d$ \cite{Fornal:2018dqn}. At the same time, in order for the vector leptoquark $X_L$ to explain the $R_{K^{(*)}}$ anomalies, one requires \cite{Fornal:2018dqn}
 \bea\label{mm1}
\frac{M_{X_L}}{g_L \sqrt{{\rm Re}\left(L_{22}^{d}L_{32}^{d*} - L_{21}^{d}L_{31}^{d*}\right)}}\approx 23 \ \rm{TeV} \ .
\eea
Because of the unitarity of the matrix $L^d$, this relation can be  fulfilled only if $M_{X_L} \lesssim(23 \ {\rm TeV })\,g_L$.
The experimental constraints  then force   $L^d$ to be of the form
\bea
L^d \,  \approx {e^{i\phi}}\left( \  
\begin{matrix} \vspace{0.5mm}
\!\delta_1 & \delta_2 & \ 1\\ \vspace{1mm}
\!e^{i\phi_1}\cos\theta&e^{i \phi_2}\sin\theta& \ \delta_3 \\ 
\!-e^{-i\phi_2} \sin\theta&e^{-i\phi_1} \cos\theta& \ \delta_4
 \end{matrix} \ \ \right) ,  
 \eea
 where $\delta_i \lesssim 0.02$. The allowed leptoquark mass in Eq.\,(\ref{mm1}) is maximized for  $\theta = \pi/4$ and $\phi_1+\phi_2 = 0$, which implies
 \bea\label{condition2}
v_L \lesssim 46 \ {\rm TeV} \ .
 \eea

 Given the assumption $v_L \gg v_\Sigma$, the lower bound on $v_\Sigma$ is of relevance. It arises from LHC dijet searches for colorons  \cite{Sirunyan:2019vgj} and translates to $v_\Sigma > 6.6 \ {\rm TeV}$. We take $v_\Sigma = 7 \ {\rm TeV}$.
The only particles other than $G'$ with masses governed by $v_\Sigma$  are the radial modes of $\hat\Sigma$ and the vector-like fermions $Q'$ and $L'$. The former do not couple to SM quarks, and our choice $v_\Sigma = 7 \ {\rm TeV}$ is consistent with experimental bounds, even for $\lambda^{(\prime)}_\Sigma$ as small as $\sim 3\times 10^{-3}$. The latter do not mix with  SM quarks, so for $Y_{ij}\sim 1$ this choice of $v_\Sigma$ is also consistent with collider searches, even for a relatively small $z$.
\vspace{1mm}
 
 In the subsequent  analysis, we consider the hierarchical symmetry breaking pattern with the following vevs,
\bea\label{hier}
&&v_R \approx 5000 \ {\rm TeV} \ , \ \ \ \ v_L \approx 40 \ {\rm TeV} \ , \ \ \ \ v_\Sigma \approx 7 \ {\rm TeV} \ . \ \ \ \ \ \ \ \ 
\eea
 The vev structure in Eq.\,(\ref{vev}) can be realized if the parameters  of the scalar potential satisfy the conditions:\break $\lambda'_{13} > 4\lambda_\Sigma (v_\Sigma/v_L)^2$, \ $\lambda'_{23} > 4\lambda_\Sigma (v_\Sigma/v_R)^2$, \ $ \lambda^{(\prime)}_\Sigma>0$ and $\kappa <0$.
We also note that the hierarchy between the vevs in Eq.\,(\ref{hier}) is not protected against  radiative corrections and\break requires a  tuning of the parameters $\lambda_{12}$, $\lambda_{13}$ and $\lambda_{23}$.

\section{Effective potential}\label{EffP}

Because of the vast  hierarchy of scales in the model and small cross terms in the scalar potential,  the three steps of symmetry breaking can be considered  independently from one  another.     Denoting the background fields as
\bea
&&\phi_R\equiv {\rm Re}(\hat\Sigma_R)_4 \sqrt{2} \ , \ \ \ \ \phi_L\equiv {\rm Re}(\hat\Sigma_L)_4 \sqrt{2} \ ,\nn\\
&&\hspace{15mm}\phi_\Sigma\equiv {\rm Re}(\hat\Sigma)^1_1 \sqrt{2} \ ,  
\eea
the effective potential  splits into three pieces,
\bea\label{pieces}
V_{\rm eff} = V_{\rm eff}^{(R)}(\phi_R)+ V_{\rm eff}^{(L)}(\phi_L)+ V_{\rm eff}^{(\Sigma)}(\phi_\Sigma) \ .
\eea
Before analyzing the phase transitions, we first discuss the effective potential in the general  case.
We adopt the  collective notation for 
the background fields $\phi = \phi_R, \phi_L, \phi_\Sigma$, the vevs $v=v_R, v_L, v_\Sigma$ and the quartic couplings $\lambda = \lambda_R, \lambda_L, \lambda_\Sigma, \lambda'_\Sigma$. Each piece of the effective potential consists of a tree-level part, a one-loop Coleman-Weinberg correction and a finite temperature contribution,
\bea
V_{\rm eff}(\phi, T) = V_{\rm tree}(\phi) + V_{\rm {loop}}(\phi)+ V_{\rm temp}(\phi, T) \ .
\eea
Using the fact that the minimum of the tree-level potential for $\phi=\phi_R,\phi_L$ is at $v = \mu/\sqrt{\lambda}$, one can write 
\bea
V_{\rm tree}(\phi) = - \frac12 \lambda \,v^2\phi^2+ \frac14 \lambda \,\phi^4   \ .
\eea
The tree-level potential for $\phi=\phi_\Sigma$ contains terms involving $\lambda_\Sigma$ and $\lambda^\prime_\Sigma$ with  different $z$-dependence. 

To obtain the  Coleman-Weinberg term, we implement the cutoff regularization scheme and assume that  the minimum of the one-loop potential and  the mass of $\phi$ are the same as their tree-level values \cite{PhysRevD.45.2685}. 
In this scheme, the one-loop zero temperature correction is 
\bea\label{fdm}
V_{\rm loop}(\phi) &=&\sum_{\rm particles}\frac{n_i}{64\pi^2} \bigg\{m_i^4(\phi)\left[\log\left(\frac{m_i^2(\phi)}{m_i^2(v)}\right)-\frac32 \right]\nn\\
&&\hspace{2mm}+ \ 2 \,m_i^2(\phi)\,m_i^2(v) \bigg\}\ , \ \ \ \ \ 
\eea
where the sum  is over all particles charged under the gauge group that undergoes symmetry  breaking, including the Goldstone bosons $\chi_{GB}$, $n_i$ is the number of degrees of freedom with an extra minus sign for fermions, and $m_i(\phi)$
are the background field-dependent masses. For the contribution of the Goldstone bosons one needs to replace $m_{\chi_{GB}}(v) \to m_\Phi(v)$, where $\Phi$ is the radial mode.

The temperature-dependent part of the potential  consists of  the one-loop finite temperature contribution $V_{\rm temp}^{(1)}(\phi,T)$ and, in case of bosonic degrees of freedom, the Daisy diagrams contribution $V_{\rm temp}^{(2)}(\phi, T)$. The corresponding formulae are given by \cite{Quiros:2007zz}
\bea
V_{\rm temp}^{(1)}(\phi, T) &=& \frac{T^4}{2\pi^2} \sum_{\rm particles}  n_i \int_0^\infty \!dy\, y^2\nn\\
&\times&  \log\left(1\mp e^{-\sqrt{{m_i^2(\phi)}/{T^2}+y^2}}\right) , \ \ \ \ \ \ \ \  
\eea
where the minus sign is for bosons and the plus sign is for fermions, and
\bea\label{debm}
V_{\rm temp}^{(2)}(\phi,  T) =\!   \frac{T}{12\pi} \sum_{\rm bosons} \!n'_i \left\{m_i^3(\phi)\!-\!\left[m_i^2(\phi)+\Pi_i(T)\right]^{\frac32}\right\}\!.\nn\\ 
\eea
The thermal  masses $\Pi_i(T)$ can be calculated following the prescription provided in \cite{Comelli:1996vm}.

\begin{figure}[t!]
\includegraphics[trim={1.8cm 0.4cm 1.8cm 0},clip,width=7.2cm]{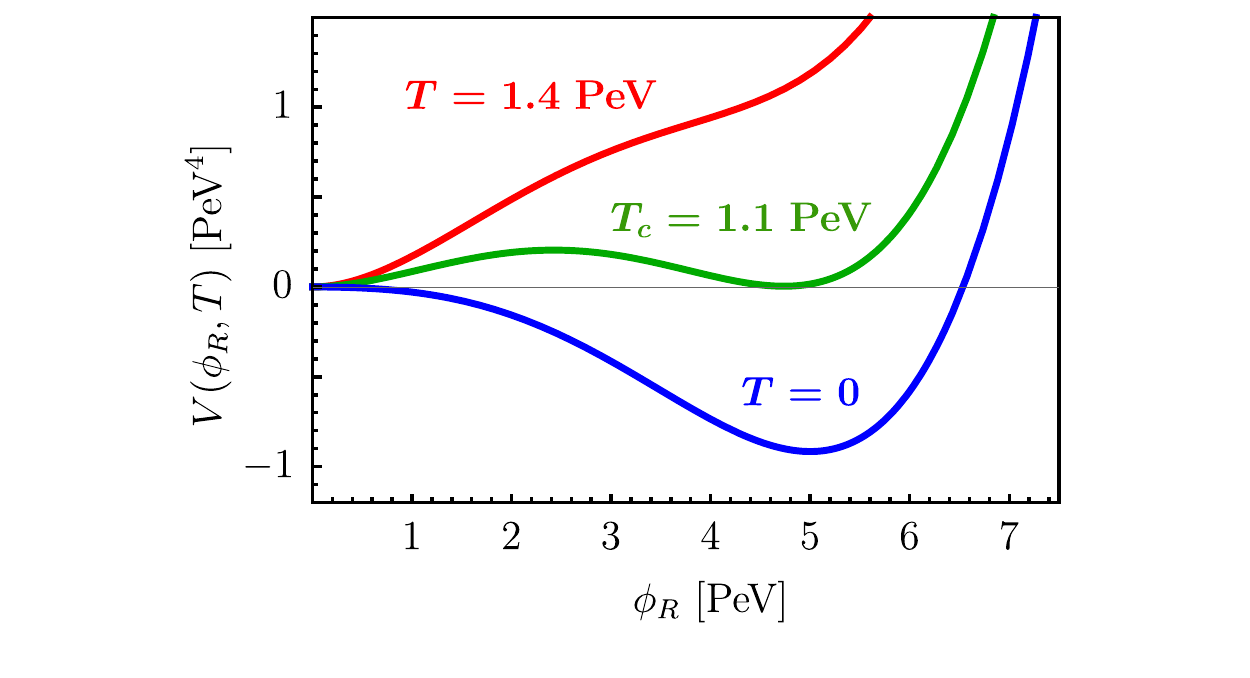}\vspace{-1mm}
\caption{The effective potential $V_{\rm eff}(\phi_R, T)$ for the temperatures:  $T=0$, \ $T_c=1.1 \ {\rm PeV}$ \  and \ $T=1.4 \ {\rm PeV}$, assuming $v_R = 5 \ {\rm PeV}$, $\ \lambda_R=0.011$ and after subtracting off  the term $V_{\rm eff}(0,T)$.}\label{fig1}\vspace{3mm}
\end{figure} 

\section{Phase transitions}\label{sec44}

A strong first order phase transition is required to produce a gravitational wave signal. This occurs when 
the effective potential develops a barrier separating the false vacuum from the true vacuum. 
We perform a scan over the parameters of the model and identify the regions of parameter space which yield gravitational wave signals most promising for detection. As a benchmark scenario, we 
adopt the values $\lambda_R(v_R) = 0.011$, $\lambda_L(v_L) = 0.029$ and $\lambda^{(\prime)}_\Sigma(v_\Sigma) = 0.036$, as this choice of parameters leads to a strong gravitational wave signal.
For such small quartics the only relevant contributions to the  field-dependent masses and thermal masses are those involving the gauge couplings $g_{R/L}$. \vspace{1mm}
  
To properly estimate those contributions, we first analyze the running of the gauge couplings. 
We match $g_R$ and $g_L$ to the SM strong coupling $g_s$ at 
 the scale $v_\Sigma = 7 \ {\rm TeV}$ via Eq.\,(\ref{six}) and choose $g_R(v_\Sigma) = g_L(v_\Sigma)$. This implies that $g_R(v_\Sigma) = g_L(v_\Sigma)\simeq1.44$. 
We then perform the running  using the renormalization group equations
\bea
\frac{\partial g_{R/L}(\mu)}{\partial \log\mu} = -\left(11-\frac{n_s}{6}-\frac{2n_f}{3}\right)\frac{g^3_{R/L}(\mu)}{16\pi^2} \ ,
\eea
where $n_s$ is the number of complex scalars and $n_f$ is the number of Dirac fermions in the fundamental representation of the gauge group ${\rm SU}(4)_R/{\rm SU}(4)_L$ with masses below the scale $\mu$. We find that $g_R(v_R) \simeq 1.01$ and $g_L(v_L) \simeq 1.32$.

\vspace{18mm}

\noindent
${(1)}$ {\underline{\emph{First phase transition}}: \ {${\rm SU}(4)_R \to {\rm SU}(3)_R$} \vspace{1mm}\\
This transition is triggered when the field $\hat{\Sigma}_R$ develops the vev as in Eq.\,(\ref{vev}) with $v_R \approx 5000 \ {\rm TeV}$. 
The relevant background field-dependent masses are
\bea\label{BFD}
&& m_{X_R}(\phi_R) = \tfrac12g_R\phi_R \ , \ \ \ \ \  m_{Z'_R}(\phi_R) =\tfrac{M_{Z'_R}}{v_R}\,{\phi_R}\ . \ \ \ \ \ 
\eea
The numbers of degrees of freedom corresponding to the gauge bosons $X_R$ and $Z'_R$ are:
$n_{X_R} \!=\! 18$ and $n_{Z'_R} \!= 3$.
The\break  thermal masses are given by
\bea\label{Deb}
&& \Pi_{X_R}^L(T) = \Pi_{Z'_R}^L(T) = \tfrac{8}{3}g_R^2  T^2  \ ,\nn\\
&& \Pi_{\Phi}(T) = \Pi_{\chi_{GB}}(T) \approx \tfrac18\,\Big(3g_R^2 +2\tfrac{M_{Z'_R}^2}{v_R^2}\Big)\,T^2 \ , \ \ \ \ 
\eea
where we dropped terms involving the small quartic coupling. The superscript $L$ for the gauge boson thermal masses denotes longitudinal components, $\Phi$ is the radial mode and $\chi_{GB}$ are the Goldstone bosons. The corresponding numbers of degrees of freedom are: $n^L_{X_R} = 6$, \,$n^L_{Z'_R} = 1$, \,$n_{\Phi} = 1$ and $n_{\chi_{GB}} = 7$.

Figure \ref{fig1} shows the full $\phi_R$-dependent part of the effective potential, $V_{\rm eff}(\phi_R,T) - V_{\rm eff}(0,T)$, for the parameter values discussed above and  for three different temperatures: $T=0$, \,$T_c=1.1 \ {\rm PeV}$ and \,$T=1.4 \ {\rm PeV}$. At the critical temperature $T_c$ the two vacua  become degenerate. The order parameter is equal to $\xi^{(R)} \equiv {(\phi_R)_c}/{T_c} \approx 4$, indicating a strong first order phase transition. 
\vspace{5mm}

\noindent
${(2)}$ {\underline{\emph{Second phase transition}}}: \ ${\rm SU}(4)_L \to {\rm SU}(3)_L$} \vspace{1mm}\\
This transition happens when the field $\hat{\Sigma}_L$ develops the vev $v_L\approx 40 \ {\rm TeV}$. 
The corresponding background field-dependent masses and thermal masses are obtained from Eqs.\,(\ref{BFD}) and (\ref{Deb})
upon substituting $R\to L$. The critical temperature is $T_c \approx 15 \ {\rm TeV}$ and the order parameter $\xi^{(L)} \approx 3$.
\vspace{2mm}

\noindent
${(3)}$ \underline{\emph{Third phase transition}}: \ {${\rm SU}(3)_R \times {\rm SU}(3)_L \to {\rm SU}(3)_c$} \vspace{1mm}\\
This symmetry breaking is triggered when $\hat\Sigma$ develops the vev as in Eq.\,(\ref{vev}) with $v_\Sigma\approx 7 \ {\rm TeV}$. For a small $z$ the contribution of the cross terms to the effective potential is small, as are those of the vector-like fermions $Q'$ and $L'$, even with Yukawas $Y_{ij}\sim 1$ (see, e.g., \cite{Angelescu:2018dkk,Davoudiasl:2012tu} for the corresponding formulae). Therefore, the only relevant background field-dependent mass is that of $G'$. For transverse components
\bea
&&m_{G'}(\phi_\Sigma) = \tfrac1{\sqrt2}\sqrt{g_R^2+g_L^2}\, \phi_\Sigma \ , 
\eea 
with the number of degrees of freedom $n_{G'}^T=16$. For the\break longitudinal modes of $G'$ and the SM gluon, the masses 
 $m_i^2(\phi_\Sigma)+\Pi_i^L(T)$
are given by the eigenvalues of the matrix
\bea
\mathcal{M}^2_i(\phi_\Sigma,T)=\frac12\begin{pmatrix}
g_R^2(\phi_\Sigma^2 + 4T^2)& -g_R g_L\phi_\Sigma^2  \\
-g_R  g_L\phi_\Sigma^2& g_L^2(\phi_\Sigma^2+4 T^2)
\end{pmatrix}  . \ \ \ \ \ \ 
\eea
The numbers of degrees of freedom are $n^L_{G'} = n^L_g=8$. The thermal masses for the radial modes and  Goldstone bosons are 
\bea
\Pi_{\Sigma}(T) \approx (g_R^2+g_L^2)\,T^2
\eea
 with $n_\Sigma = 32$. 
A strong first order phase transition occurs since $\xi^{(\Sigma)}\approx2$ for the critical temperature $T_c \approx 1.7 \ {\rm TeV}$.

\section{Gravitational wave signals}

As a result of a first order phase transition, bubbles of true vacuum are nucleated, they expand (with velocity $v_w$) and eventually fill up the entire universe. The bubble nucleation rate per unit volume is given by the expression \cite{LINDE1983421}
\bea
\Gamma(T) \approx \left(\frac{S_3(T)}{2\pi T}\right)^{\frac32}T^4\,e^{-\frac{S_3(T)}{T}} \ ,
\eea
where $S_3(T)$ is the Euclidean action 
\bea
S_3(T)= 4\pi \int dr\,r^2\left[\frac12\left(\frac{d\phi_b}{dr}\right)^2+V_{\rm eff}(\phi_b, T)\right]. \ \ \ \  \
\eea
Here $\phi_b(r)$ is the ${\rm SO}(3)$ symmetric bounce solution describing the profile of the expanding bubble, i.e., the solution of  the equation 
\bea
\frac{d^2 \phi_b}{dr^2}+\frac{2}{r}\frac{d\phi_b}{dr}-\frac{dV_{\rm eff}(\phi,T)}{d\phi}\bigg|_{\phi=\phi_b} = 0
\eea
with the boundary conditions
\bea
\frac{d\phi_b}{dr}\bigg|_{r=0} = 0 \ , \ \ \ \ \ \phi_b(\infty) = \phi_{\rm false} \ ,
\eea
where $\phi_{\rm false}=0$ is the field value of the false vacuum. 

The phase transition begins at the temperature $T_*$, called the nucleation temperature,  at which  $\Gamma(T_*) \approx H^4$, where $H$ is the Hubble value at that time. This is equivalent to 
\bea\label{ST4}
\frac{S_3(T_*)}{T_*}   \approx  4\log\left(\frac{M_{P}}{T_*}\right)  - \log\left[\left(\frac{4\pi^3g_*}{45}\right)^{\!\!2}\!\left(\frac{2\pi \,T_*}{S_3(T_*)}\right)^{\!\!\frac32}\right],\nn\\
\eea
where $M_{P} = 1.22\times 10^{19} \ {\rm GeV}$ is the Planck mass. \vspace{1mm}

Each phase transition continues until most of the universe is filled with bubbles of true vacuum. 
The inverse of the duration of this process, the so-called $\tilde\beta$ parameter, is given by
\bea
\tilde{\beta} \equiv T_* \frac{d}{dT} \!\left(\frac{S_3(T)}{T}\right)\bigg|_{T=T_*} \ .
\eea
The strength of a phase transition, denoted by $\alpha$, is defined as
\bea
\alpha \equiv \frac{\rho_{\rm vac}(T_*)}{\rho_{\rm rad}(T_*)} \ ,
\eea
where $\rho_{\rm vac}(T_*)$ is the energy density of the false vacuum (i.e., the latent heat released during the phase transition) \cite{Ellis:2018mja},
\bea
\rho_{\rm vac}(T_*) =\Delta V_{\rm eff}(T_*)-T_*\frac{\partial \Delta V_{\rm eff}(T)}{\partial T}\bigg|_{T=T_*}
\eea
with  \ \
$
\Delta V_{\rm eff}(T) = V_{\rm eff}(\phi_{\rm false}, T)- V_{\rm eff}(\phi_{\rm true}, T) \ ,
$\\[5pt] 
and ${\rho_{\rm rad}(T_*)}$ is the energy density of radiation at nucleation temperature,
\vspace{-7mm}

\bea
\rho_{\rm rad}(T_*) = \frac{\pi^2}{30} g_* T^4_* \ .
\eea

\noindent
In the expressions above $\phi_{\rm true}$ is the field value of the true vacuum, whereas 
$g_*$ is the number of relativistic degrees of freedom at the time of the transition. In our benchmark\break scenario: $g_*^{(1)}\simeq274$, \,$g_*^{(2)}\simeq252$ \,and\, $g_*^{(3)}\simeq228$.
The four  parameters: $\alpha$, $\tilde\beta$, $v_w$ and $T_*$ determine the size and peak frequency of the 
 stochastic gravitational wave signal. In our analysis we set the bubble wall velocity to 
$v_w = 0.6\,c$ (for an extensive discussion of bubble expansion, see \cite{Espinosa:2010hh,Caprini:2015zlo}).
\vspace{1mm}

\begin{figure}[t!]
\includegraphics[trim={1.45cm 0.4cm 1.9cm 0},clip,width=8cm]{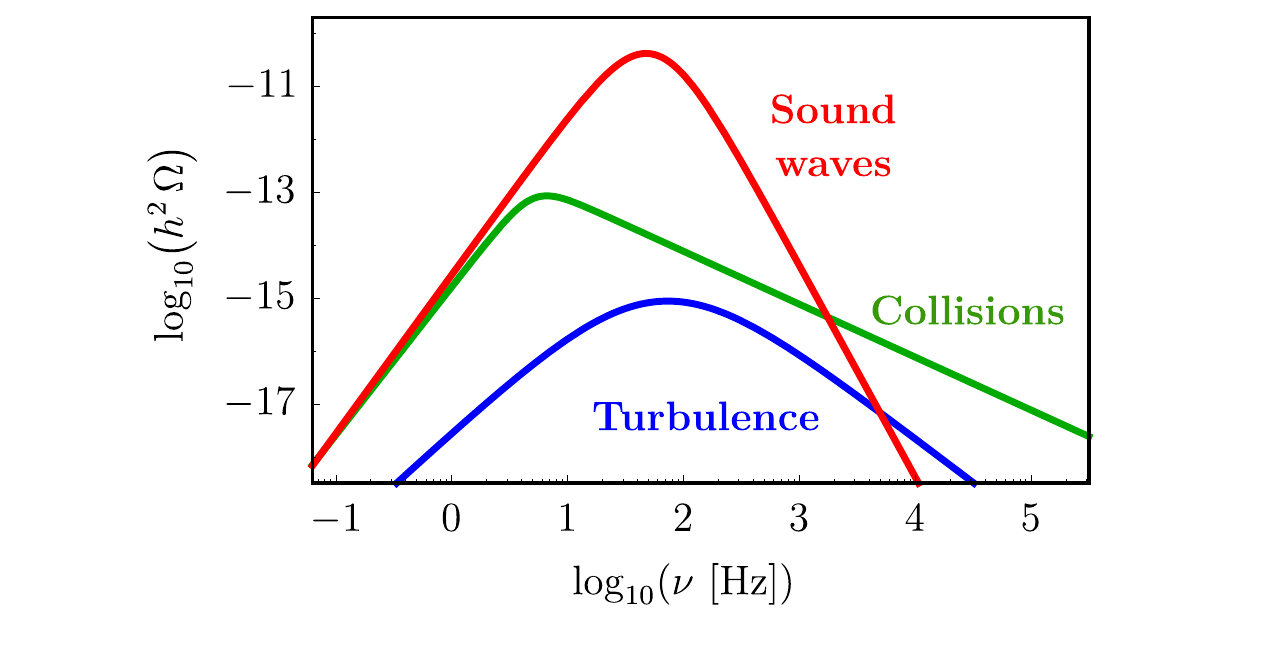}\vspace{-1mm}
\caption{Contributions to the gravitational wave signal of the phase transition  ${\rm SU}(4)_R \to {\rm SU}(3)_R$ arising from sound waves, bubble collisions and magnetohydrodynamic turbulence.\vspace{1.2mm}}\label{fig2}
\end{figure} 

There are three sources of gravitational waves generated  from phase transitions: sound waves, bubble collisions and magnetohydrodynamic turbulence. Those three contributions combine linearly to give the total gravitational wave signal
\bea\label{alll}
h^2 \Omega_{\rm GW} \approx  h^2 \Omega_{\rm sound} + h^2 \Omega_{\rm collision}+ h^2 \Omega_{\rm turbulence}  \ . \ \ \ \ \ 
\eea 
\vspace{-1mm}

\noindent
The contribution from sound waves is \cite{Hindmarsh:2013xza,Caprini:2015zlo}
\bea\label{sound}
 h^2\Omega_{\rm sound}(\nu) &\ \approx \ & (1.86\times 10^{-5})\  \frac{v_w}{\tilde\beta}\left(\frac{\kappa_s\, \alpha}{1+\alpha}\right)^2\left(\frac{100}{g_*}\right)^{\frac13} \nn\\[-1pt]
 &\times& \frac{\big(\frac{\nu}{\nu_s}\big)^{3}}{\Big[1+0.75\,\big(\frac{\nu}{\nu_s}\big)^{2}\,\Big]^{\frac72}} \ ,
\eea
where the model-dependent parameter $\kappa_s$ is the fraction of the latent heat that is transformed into the bulk motion of the plasma, approximated by \cite{Espinosa:2010hh}
\bea
\kappa_s\approx \frac{\alpha}{0.73+0.083\sqrt\alpha+\alpha} \ ,
\eea
and $\nu_s$ is the peak frequency given by
\bea\label{sound2}
\nu_s &=& (0.019 \ {\rm Hz} )\, \frac{\tilde\beta}{v_w}\,\left(\frac{g_*}{100}\right)^\frac16\left(\frac{T_*}{100 \ {\rm TeV}}\right) \ .\ \ \ \ \ \ \ 
\eea
\vspace{0mm}

\noindent
The contribution from bubble collisions is \cite{Kosowsky:1991ua,Huber:2008hg,Caprini:2015zlo}
\bea
h^2 \Omega_{\rm collision}(\nu) &\ \approx \ & (1.66\times 10^{-5})\,  \frac{1}{\tilde\beta^2} \left(\frac{\kappa_c \,\alpha}{1+\alpha}\right)^2\left(\frac{100}{g_*}\right)^{\frac13}\nn\\
 &\times&\left(\frac{v_w^3}{1+2.4\,v_w^2}\right)\frac{\big(\frac{\nu}{\nu_c}\big)^{2.8}}{1+2.8\big(\frac{\nu}{\nu_c}\big)^{3.8}} \ ,
\eea
where $\kappa_c$ is  the fraction of the latent heat that is deposited into a thin shell close to the bubble front \cite{Kamionkowski:1993fg},
\bea
\kappa_c \approx  \frac{0.715\,\alpha+\frac{4}{27}\sqrt{{\frac{3\alpha}{2}}}}{1+0.715\,\alpha}\ ,
\eea
and the peak frequency $\nu_c$ is
\bea
\nu_c &=& (0.010 \ {\rm Hz} )\, \tilde\beta\,\left(\frac{g_{*}}{100}\right)^\frac16\left(\frac{T_*}{100 \ {\rm TeV}}\right)\nn\\[-2pt]
&\times& \left(\frac{1}{1.8-0.1v_w+v_w^2}\right) \ .
\eea
\vspace{1mm}

\begin{figure}[t!]
\includegraphics[trim={0cm 0.0cm 0cm 0},clip,width=8.1cm]{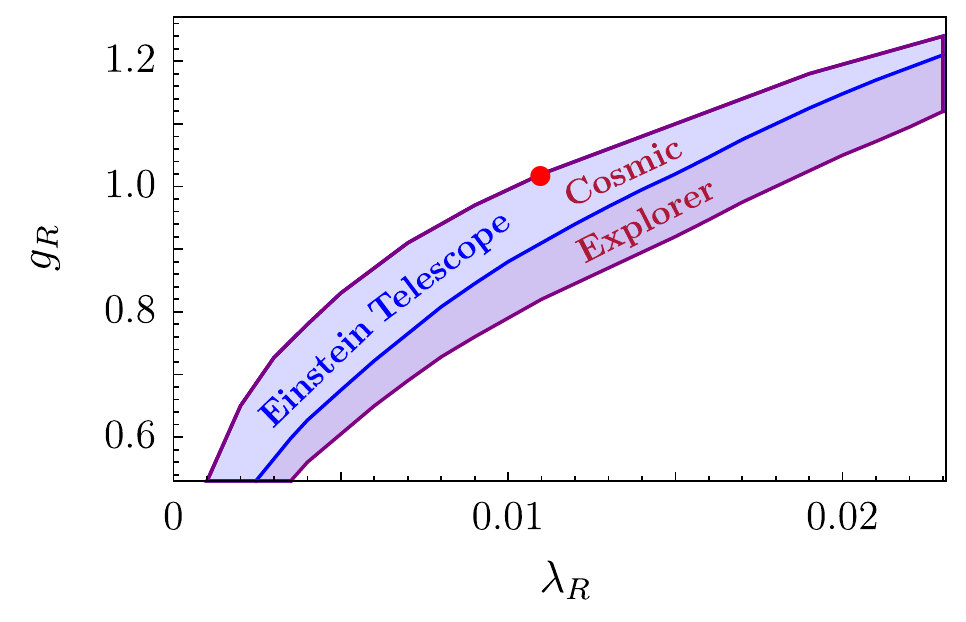}\vspace{-1.0mm}
\caption{Regions of parameter space for which a phase transition  ${\rm SU}(4)_R \to {\rm SU}(3)_R$  with $v_R = 5 \ {\rm PeV}$ gives rise to a gravitational wave signal detectable at the Einstein Telescope (blue) and  Cosmic Explorer (blue and purple) with a signal-to-noise ratio greater than five after one year of collecting data (see text for details). The red dot denotes the benchmark parameters  adopted  in Fig.\,\ref{fig4}.\vspace{0.5mm}}\label{fig3}
\end{figure} 

\noindent
The contribution from turbulence is \cite{Caprini:2006jb,Caprini:2009yp}
\bea
h^2 \Omega_{\rm turbulence}(\nu) &\ \approx \ & (3.35\times 10^{-4}) \, \frac{v_w}{\tilde\beta}\!\left(\frac{\kappa_t \,\alpha}{1+\alpha}\right)^{\!\frac32}\!\left(\frac{100}{g_*}\right)^{\!\frac13}\nn\\[-1pt]
 &\times& \frac{\big(\frac{\nu}{\nu_t}\big)^{3}}{\big(1+\frac{8\pi \nu}{h_*}\big)\big(1+\frac{\nu}{\nu_t}\big)^{\frac{11}{3}}} \ ,
\eea
where $\kappa_t = \epsilon\,\kappa_s$ denotes the fraction of the latent heat transformed into magnetohydrodynamic turbulence  (following \cite{Caprini:2015zlo}, we take $\epsilon=0.05$), 
the peak frequency $\nu_t$ is
\bea
\nu_t &=& (0.027 \ {\rm Hz} )\, \frac{\tilde\beta}{v_w}\,\left(\frac{g_*}{100}\right)^\frac16\left(\frac{T_*}{100 \ {\rm TeV}}\right) \ \ \ \ \ 
\eea
and the parameter $h_*$ \cite{Caprini:2015zlo}
\bea\label{hstar}
h_* = (0.0165 \ {\rm Hz})\left(\frac{g_*}{100}\right)^\frac16\left(\frac{T_*}{100 \ {\rm TeV}}\right) \ .
\eea

\begin{figure*}[t!]
\includegraphics[trim={0cm 0.0cm 0 0},clip,width=15cm]{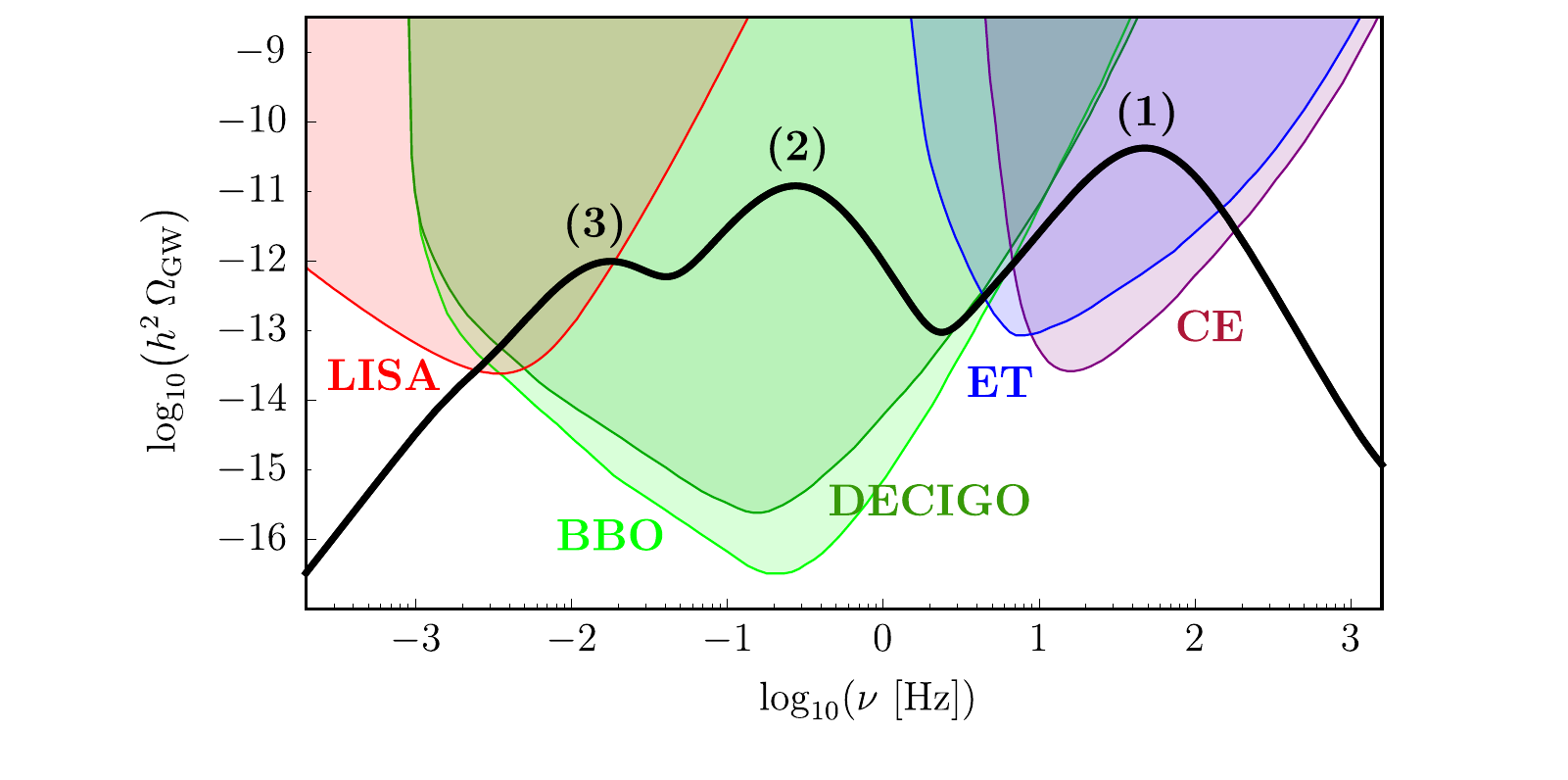} \vspace{-3mm}
\caption{Gravitational wave signature of the Left-Right SU(4) Model (black line) for the benchmark scenario described in Eq.\,(\ref{hier}). Overplotted are the sensitivities of the future gravitational wave experiments: LISA in the C1 configuration  \cite{Caprini:2015zlo} (red), Big Bang Observer \cite{Yagi:2011wg} (light green), DECIGO \cite{Yagi:2011wg} (dark green), Einstein Telescope \cite{Sathyaprakash:2012jk} (blue) and Cosmic Explorer \cite{Reitze:2019iox} (purple). The three peaks correspond to the phase transitions: \ {\rm{$(1)\,$}} {${\rm SU}(4)_R \to {\rm SU}(3)_R$}, \ {\rm{$(2)\,$}}  {${\rm SU}(4)_L \to {\rm SU}(3)_L$} \,and\,  {\rm{$(3)\,$}}  {${\rm SU}(3)_R \times {\rm SU}(3)_L \to {\rm SU}(3)_c$} discussed in Sec.\,\ref{sec44}.\vspace{6mm}}\label{fig4}
\end{figure*}

To find the gravitational wave signal from  the three phase transitions, we analyzed separately the $\phi_R$, $\phi_L$, $\phi_\Sigma$-dependent pieces of the potential in Eq.\,(\ref{pieces}). In each case we determined numerically  the temperature at which the shape of the potential yields the Euclidean action $S(T_*)$ satisfying Eq.\,(\ref{ST4}). For this nucleation temperature, we calculated the values of the parameters $\alpha$, $\tilde\beta$, and used them to derive the gravitational wave spectrum via  Eqs.\,(\ref{alll})-(\ref{hstar}).
\newpage

Our calculation revealed that in the Left-Right SU(4) Model the sound wave contribution dominates over the contributions from bubble collisions and magnetohydrodynamic turbulence in most of the peak region, thus the shape of the signal is well approximated by Eq.\,(\ref{sound}) and the peak frequency by Eq.\,(\ref{sound2}). This is illustrated in Fig.\,\ref{fig2} for the   phase transition associated with ${\rm SU}(4)_R \to {\rm SU}(3)_R$.

In order to assess how generic are the phase transitions leading to detectable gravitational wave signals  in the model, we performed a scan over the relevant parameters. Figure \ref{fig3}  presents the regions of parameter space for the gauge coupling $g_R$ versus the quartic coupling $\lambda_R$ which yield  a first order phase transition  ${\rm SU}(4)_R \to {\rm SU}(3)_R$  (for $v_R = 5 \ {\rm PeV}$) giving rise to signals detectable at the Einstein Telescope and Cosmic Explorer, with a signal-to-noise ratio greater than five after one year of collecting data. In particular, the lower boundaries of those regions correspond to the signal-to-noise ratio of five for each experiment. Above the upper boundary,  either the value of $S_3(T)/T$ is too large to satisfy the condition in Eq.\,(34) or  the zero temperature vacuum at $v_R \ne 0$ has a  higher energy density than the vacuum at $v_R=0$, which is  unphysical, since it would lead to a second transition back to the vacuum with $v_R=0$.
Similarly sized regions of parameter space relevant for the phase transitions  ${\rm SU}(4)_L \to {\rm SU}(3)_L$ (for $v_L = 40 \ {\rm TeV}$) and  ${\rm SU}(3)_R \times {\rm SU}(3)_L \to {\rm SU}(3)_c$ (for $v_\Sigma = 7 \ {\rm TeV}$) yield signals detectable in other gravitational wave experiments.

Figure \ref{fig4} shows the combined  spectrum of gravitational waves from all three phase transitions in our benchmark\break scenario, with the corresponding parameters summarized in Table \ref{tab33}.
As expected, each of the phase transitions produces a distinct peak in the spectrum, characterized by a large signal-to-noise ratio. As seen from Eq.\,(\ref{sound2}), the position of  individual peaks depends linearly on the nucleation temperature $T_*$, thus signals from phase transitions corresponding to higher symmetry breaking scales appear at higher frequencies. The peak frequency  depends  also linearly on $\tilde\beta$. 
The height of the peak is governed by $\alpha$ and $\tilde\beta$\,; it increases with bigger $\alpha$ and decreases with larger $\tilde\beta$.
 Of course all those parameters depend on the values of the vevs, quartic couplings and gauge couplings in the model. 

\begin{table}[t!] 
\begin{center}
\begingroup
\setlength{\tabcolsep}{6pt} 
\renewcommand{\arraystretch}{1.5} 
\begin{tabular}{  |c |c |c| c|} 
\hline
Phase transition & $\alpha$ & $\tilde\beta$   & $T_*$ \\ 
\hline
\hline
$\hspace{-13.5mm}(1) \ \ \ {\rm SU}(4)_R \to {\rm SU}(3)_R$ & 0.35 & 270 & 430  \ {\rm TeV} \\[1pt]
\hline
$\hspace{-13.5mm}(2) \ \ \ {\rm SU}(4)_L \to \,{\rm SU}(3)_L$ & 0.09 & 120 &  6.4  \ {\rm TeV} \\[1pt] 
\hline
$(3) \ \ \ {\rm SU}(3)_R \times {\rm SU}(3)_L \to {\rm SU}(3)_c$& 0.02 & 60 &  0.9  \ {\rm TeV} \\[1pt]
\hline
\end{tabular}
\endgroup
\end{center}
\vspace{-1mm}
\caption{Values of the parameters $\alpha$, $\tilde\beta$ and $T_*$ for the three phase transitions giving rise to the gravitational wave signal  in Fig.\,\ref{fig4}.}\vspace{1mm}
\label{tab33}
\end{table}

Within the benchmark scenario, the gravitational wave signal generated by the symmetry breaking ${\rm SU}(4)_R \to {\rm SU}(3)_R$ at the scale $v_R \approx 5 \ {\rm PeV}$ (peak $(1)$) falls within the sensitivity of the future gravitational wave detectors Cosmic Explorer and Einstein Telescope. The signal resulting from the second phase transition ${\rm SU}(4)_L \to {\rm SU}(3)_L$ at the scale $v_L \approx 40 \ {\rm TeV}$ (peak $(2)$) is well within the reach of the Big Bang Observer and  DECIGO. The third phase transition ${\rm SU}(3)_R \times {\rm SU}(3)_L \to {\rm SU}(3)_c$ occurring at the scale $v_\Sigma \approx 7 \ {\rm TeV}$ (peak $(3)$) can also be probed by the Big Bang Observer and DECIGO. In\break addition, it can be searched for by LISA, but only if the C1 configuration \cite{Caprini:2015zlo} is implemented.

A unique property of the Left-Right SU(4) Model is that the range of peak frequencies  for the  phase transitions  ${\rm SU}(4)_L \to {\rm SU}(3)_L$ and ${\rm SU}(3)_R \times {\rm SU}(3)_L \to {\rm SU}(3)_c$
is constrained by the size of the $R_{K^{(*)}}$ anomalies, as described by Eq.\,(\ref{condition2}). In our benchmark scenario we assumed that there is a maximal hierarchy between the scales $v_L$ and $v_\Sigma$, which leads to  two well-separated peaks in the spectrum. However, if the two scales are  comparable, then the size of the flavor anomalies sets the symmetry breaking scale  at $v_L \approx v_\Sigma \lesssim 25 \ {\rm TeV}$, resulting  in a single peak shifted towards lower frequencies compared to peak $(2)$ in Fig.\,\ref{fig4}. This is still within the reach of the Big Bang Observer and DECIGO.
\vspace{1mm}

Finally, we point out that the scale of the  symmetry breaking ${\rm SU}(4)_R \to {\rm SU}(3)_R$  is not bounded from above. In particular, it can be larger than $\sim 100 \ {\rm PeV}$, shifting peak $(1)$ to higher frequencies and escaping the detection at the Cosmic Explorer and Einstein Telescope. A  gravitational wave experiment sensitive to such high frequencies would be necessary to probe this scenario.

\section{Conclusions}

Gravitational wave experiments have recently emerged as a powerful tool for testing particle physics models. One class of signatures which those experiments are sensitive to arises from first order phase transitions in the early universe, making them valuable probes of the scalar sector in models with spontaneous symmetry breaking. 

\vspace{25mm}

In this paper we demonstrated, in the context of the Left-Right SU(4) Model,  that  gravitational wave detectors can be used to look for signatures specific to the flavor anomalies recently observed at the LHCb, BaBar and Belle experiments. The measured magnitude of lepton universality violation  implies that there can be two peaks in the gravitational wave spectrum within the sensitivity of the upcoming  LISA, Big Bang Observer and DECIGO experiments. There is also a possibility of a  third peak which could be observed by the Cosmic Explorer and Einstein Telescope.
\vspace{3mm}

If the hints of lepton universality violation are confirmed and a gravitational wave signal with features similar to those of the Left-Right SU(4) Model is discovered, this would  be a strong motivation for building the 100 TeV collider, which could provide a complementary direct detection method of testing the model.

\vspace{20mm}

\subsection*{Acknowledgments}
The author is grateful to Peter Stangl and Yue Zhao for\break inspiring discussions and helpful comments.
This research was supported in part by the U.S. Department of Energy\break under Award
No. ${\rm DE}$-${\rm SC0009959}$.

\bibliography{refs}

\end{document}